\documentstyle[aps,preprint]{revtex}

\begin{document}
\draft \tightenlines \preprint{cond-mat/9912472}

\title{Nonequilibrium Quantum Evolution of Open Systems
\footnote{Contributed to the journal "Condensed Matter Physics".}}

\author{Sang Pyo Kim}
\address{Department of Physics\\ Kunsan National University\\ Kunsan 573-701,
Korea}

\date{\today}

\maketitle

\begin{abstract}
We apply the Liouville-von Neumann (LvN) approach to open systems
to describe the nonequilibrium quantum evolution. The
Liouville-von Neumann approach is a unified method that can be
applied to both time-independent (closed) and time-dependent
(open) systems and to both equilibrium and nonequilibrium systems.
We study the nonequilibrium quantum evolution of oscillator models
for open boson and fermion systems

\vspace{1cm}


\vspace{.5cm}


\end{abstract}


\section{Introduction}

The system of an open system interacts with an environment and
depends on time implicitly or explicitly. A closed system can
depend on time only when it is placed in an expanding Universe. It
is the parametric interaction that forces the coupling constants
of the system to depend on time. The time-dependent system can not
maintain the equilibrium state since the density operator defined
in terms of the Hamiltonian does not satisfies the Liouville-von
Neumann equation. Hence, the open system requires the
nonequilibrium evolution through the interaction with the
environment. Recently the nonequilibrium evolution of open quantum
systems has been an issue of much concern and debate
\cite{lindenberg}.

There have been two widely used methods to treat the
nonequilibrium quantum systems. One is time-path integral method
where path integrals are carried out over a contour of complex
(imaginary) time to take into account the thermal nature of the
system (for review and references, see \cite{landsman}). The other
is thermal field dynamics where the system of the open system
interacts and makes a closed system with a fictitious system (for
review and references, see \cite{umezawa}).

It is the purpose of this paper to apply a recently developed
field theoretical method, the so-called the Liouville-von Neumann
(LvN) approach to nonequilibrium quantum systems (for references,
see \cite{kim1}). It is a canonical method where the evolution of
open systems is governed by the Schwinger-Tomonaga equation or the
functional Schr\"{o}dinger equation. The LvN approach is well
suited to describe the nonequilibrium quantum processes because
the same LvN equation is used both to construct the Hilbert space
of quantum states and to find the density operators. Also the
method is nonperturbative in that the lowest approximation is the
time-dependent Gaussian wave functional for a self-interacting
system.

The organization of this paper is as follows. In Sec. II we review
the LvN approach and apply to a time-dependent oscillator. In Sec.
III we study three oscillator models for open boson systems and
find their nonequilibrium evolution. In Sec. IV we study open
fermion systems motivated by a Dirac field interacting with an
external source and a scalar field. In Sec. V we conclude the
paper.

\section{Liouville-von Neumann (LvN) Approach}

The LvN approach is based on the idea by Lewis and Riesenfeld that
the LvN equation can be used to find the exact quantum states of a
time-dependent quantum system \cite{lewis}. In particular, they
solved exactly the time-dependent quantum oscillator in terms of
an auxiliary equation. Recently, the author has found an
oscillator representation, where the Fock space is constructed in
terms of the classical solution \cite{kim1,kim2,kim3,kim4,kim5}.
Furthermore, the selection rule was put forth for the vacuum state
for the time-dependent oscillator, which satisfies all criteria
frequently imposed for well-known systems \cite{kim5}.

We elucidate the LvN approach to the nonequilibrium quantum
evolution of open quantum systems. The LvN approach is based on
two assumptions which are physically well-grounded. The one is
that an open system should be described by a time-dependent
quantum system obeying the Schr\"{o}dinger equation ($\hbar$ = 1)
\begin{equation}
i \frac{\partial}{\partial t} \vert \Psi(t) \rangle = \hat{H} (t)
\vert \Psi (t) \rangle. \label{sch eq}
\end{equation}
The other is that the statistical nature of an ensemble of the
system should be described by the density operator satisfying the
quantum LvN equation. In methodology the LvN approach is based on
the the following two theorems.

{\it Theorem I}. If $\hat{\cal O} (t)$ is a time-dependent
invariant operator satisfying the quantum LvN equation
\begin{equation}
i \frac{d}{dt} \hat{\cal O} (t) = i \frac{\partial}{\partial t}
\hat{{\cal O}} (t) + [ \hat{{\cal O}} (t), \hat{H} (t)] = 0,
\label{ln eq}
\end{equation}
then its eigenstates
\begin{equation}
\hat{\cal O} (t) \vert \lambda_n, t \rangle = \lambda_{n} \vert
\lambda_n, t \rangle
\end{equation}
provide the exact quantum state for Eq. (\ref{sch eq}) by
\begin{equation}
\vert \Psi (t) \rangle = \sum_{n} c_n \exp \Bigl[i \int dt \langle
\lambda_n, t \vert i \frac{\partial}{\partial t} - \hat{H} (t)
\vert \lambda_n, t \rangle \Bigr] \vert \lambda_n, t \rangle.
\label{lin sup}
\end{equation}
This theorem was first proved by Lewis and Riesenfeld. In order to
prove the theorem we use Eqs. (\ref{sch eq}) and (\ref{ln eq}) to
get the identity
\begin{equation}
 \Bigl(\hat{\cal O} (t) - \lambda_n \hat{I} \Bigr) \Bigl[ \Bigl( i
\frac{\partial}{\partial t} - \hat{H} (t) \Bigr) \vert \lambda_n,
t \rangle \Bigr] = 0. \label{ker eq}
\end{equation}
Then Eq. (\ref{ker eq}) implies that
\begin{equation}
 \Bigl(i \frac{\partial}{\partial t} - \hat{H} (t) \Bigr)
\vert \lambda_n, t \rangle = \alpha_n (t) \vert \lambda_n, t
\rangle.
\end{equation}

{\it Theorem II}. If $\hat{\cal O}_1 (t)$ and $\hat{\cal O}_2 (t)$
satisfy the LvN equation (\ref{ln eq}), then $\hat{\cal O}_1 (t)
\hat{\cal O}_2 (t)$ satisfies the same LvN equation. This follows
from the identity
\begin{eqnarray}
&&i \frac{\partial}{\partial t} \Bigl( \hat{\cal O}_1 (t)
\hat{\cal O}_2 (t) \Bigr) + \Bigl[ \Bigl( \hat{\cal O}_1 (t)
\hat{\cal O}_2 (t) \Bigr), \hat{H} (t) \Bigr]  = \nonumber\\
&&\Bigl\{ i \frac{\partial}{\partial t} \hat{\cal O}_1 (t) +
\Bigl[\hat{\cal O}_1 (t) , \hat{H} (t) \Bigr] \Bigr\} \hat{\cal
O}_2 (t) + \hat{\cal O}_1 (t) \Bigl\{ i \frac{\partial}{\partial
t} \hat{\cal O}_2 (t) + \Bigl[\hat{\cal O}_2 (t) , \hat{H} (t)
\Bigr] \Bigr\} = 0.
\end{eqnarray}
In general, one can show that $F[\hat{\cal O}(t)]$ satisfies the
LvN equation, if it is an analytical functional of $\hat{\cal O}
(t)$ and $\hat{\cal O} (t)$ satisfies the LvN equation.

A few remarks should be made. Firstly, the LvN approach can also
be applied to time-independent systems without any modification.
In this case one can choose $\hat{\cal O} = \hat{H}$ and the
energy eigenstates are the exact quantum states as expected from
quantum mechanics. Secondly, according to the theorem II any
invariant operator $\hat{\cal O} (t)$ can be used to define a
density operator
\begin{equation}
\hat{\rho}_{\cal O} (t) = \frac{1}{{\rm Tr} \Bigl(e^{- \beta
\hat{\cal O} (t)} \Bigr)}e^{- \beta \hat{\cal O} (t)}.
\end{equation}
To describe the initial thermal equilibrium state a particular
kind of invariant operator will be selected. In this sense the LvN
approach can be applied to the non-equilibrium quantum evolution
of time-dependent systems. Finally, the LvN approach can be
extended to quantum fields whose coupling constants depend on
time. When a field is appropriately mode-decomposed, the system of
the quantum field becomes an infinite system of finite-dimensional
quantum mechanical systems.

We now turn to the most typical system of time-dependent
oscillators. These systems occur in all areas of physics and play
a prominent role partly because any system around an extremum can
be approximated by a system of oscillators and partly because the
classical and quantum evolution can be found exactly. Even a free
field can be treated as a system of infinite number of oscillators
via mode-decomposition. Thus the study of a time-dependent
oscillator will shed light on the nonequilibrium quantum evolution
of a generic time-dependent quantum system.

Without loss of generality, we consider an oscillator whose mass
and frequency depend on time explicitly:
\begin{equation}
\hat{H} (t) = \frac{1}{2m(t)} \hat{p}^2 + \frac{m(t)}{2}
\omega^2(t) \hat{q}^2.
\end{equation}
In Refs. \cite{kim4,kim5} a pair of invariant operators are found
\begin{eqnarray}
\hat{A} (t) &=&  i \Bigl(u^* (t) \hat{p} - m(t) \dot{u}^*
(t)\hat{q} \Bigr), \nonumber\\ \hat{A}^{\dagger} (t) &=&  - i
\Bigl(u (t) \hat{p} - m(t) \dot{u} (t) \hat{q} \Bigr),
\label{an-cr}
\end{eqnarray}
where $u$ satisfies the classical equation of motion
\begin{equation}
\ddot{u} (t) + \frac{\dot{m}(t)}{m(t)} \dot{u} (t) + \omega^2 (t)
u(t) = 0.
\end{equation}
These invariant operators can be made the annihilation and
creation operators by imposing the standard commutation relation
for all times
\begin{equation}
[\hat{A} (t), \hat{A}^{\dagger} (t)] = 1,
\end{equation}
which is guaranteed by the wronskian condition
\begin{equation}
m(t) \Bigl(\dot{u}^* (t) u(t) - \dot{u} (t) u^* (t) \Bigr) = i.
\end{equation}

From the theorem II we can find a particularly simple functional
of two invariant operators (\ref{an-cr}) called the number
operator
\begin{equation}
\hat{N} (t) = \hat{A}^{\dagger} (t) \hat{A} (t).
\end{equation}
The vacuum state and the number states are defined by
\begin{eqnarray}
\hat{A} (t) \vert 0, t \rangle = 0, \nonumber\\ \vert n, t \rangle
= \frac{1}{\sqrt{n!}} \Bigl(\hat{A}^{\dagger} (t) \Bigr) \vert 0,
t \rangle. \label{num st}
\end{eqnarray}
According to the theorem I the exact quantum state (\ref{lin sup})
is a linear superposition of number states. The density operator
is defined by
\begin{equation}
\hat{\rho} (t) = \frac{1}{Z} e^{- \beta \omega_0 \hat{N} (t)},
\label{den op}
\end{equation}
where $Z$ is the partition function
\begin{eqnarray}
Z &=& \sum_{n = 0}^{\infty} \langle n, t \vert e^{- \beta \omega_0
\hat{N}(t)} \vert n, t \rangle \nonumber\\ &=& \frac{1}{1 -
e^{-\beta \omega_0}}.
\end{eqnarray}
The density operator has the coordinate representation
\begin{eqnarray}
\rho (x, x', t) &=& \frac{1}{Z} \sum_{n, n' = 0}^{\infty} \langle
x \vert n, t \rangle \langle n, t \vert e^{- \beta \omega_0
\hat{A}^{\dagger} (t) \hat{A} (t)} \vert n', t \rangle \langle n',
t \vert x \rangle \nonumber\\ &=& \frac{1}{Z} \sum_{n =
0}^{\infty} \Psi_n (x, t) \Psi^*_n (x', t) e^{- \beta \omega_0 n},
\end{eqnarray}
where
\begin{equation}
\Psi_{n} (x, t) = \langle x \vert n, t \rangle
\end{equation}
are the harmonic oscillator wave functions.

\section{Open Boson System}

In this section we shall consider three oscillator models for open
boson systems. These systems are motivated by a scalar field in a
linear or quadratic interaction with an environment. The first two
models have all the essential characteristics of a scalar field
with a quadratic interaction with an external source
\begin{equation}
{\cal L} = \frac{1}{2} \partial_{\mu} \phi({\bf x}, t)
\partial^{\mu} \phi({\bf x}, t) - \frac{m^2_0}{2} \phi^2 ({\bf x},
t) + j ({\bf x}, t) \phi^2 ({\bf x}, t).
\end{equation}
The last model mimics the scalar field interacting linearly with
the external source
\begin{equation}
{\cal L} = \frac{1}{2} \partial_{\mu} \phi({\bf x}, t)
\partial^{\mu} \phi({\bf x}, t) - \frac{m^2_0}{2} \phi^2 ({\bf x},
t) + j ({\bf x}, t) \phi ({\bf x}, t).
\end{equation}

As the simplest open system we consider a time-dependent
oscillator described by the single-mode Hamiltonian
\begin{equation}
\hat{H} (t) = \Omega^{D} (t) \Bigl(\hat{a}^{\dagger} \hat{a} +
\frac{1}{2} \Bigr) + \Omega^{(O)} (t) \Bigl(\frac{1}{2}
\hat{a}^{\dagger 2} + \frac{1}{2} \hat{a}^2 \Bigr), \label{os1}
\end{equation}
where
\begin{eqnarray}
\hat{a} &=& \sqrt{\frac{\omega_0}{2}} \hat{q} + i \frac{1}{\sqrt{2
\omega_0}} \hat{p}, \nonumber\\ \hat{a} &=&
\sqrt{\frac{\omega_0}{2}} \hat{q} - i \frac{1}{\sqrt{2 \omega_0}}
\hat{p}.
\end{eqnarray}
These operators are the standard annihilation and creation
operators of the time-independent oscillator
\begin{equation}
\hat{H}_0 = \frac{1}{2} \hat{p}^2 + \frac{\omega_0^2}{2}
\hat{q}^2.
\end{equation}
We look for a pair of first order invariant operators of the form
\begin{eqnarray}
\pmatrix{ \hat{A} (t) \cr \hat{A}^{\dagger} (t)} = \pmatrix{u (t)
& v(t) \cr v^*(t) & u^* (t)} \pmatrix{\hat{a} \cr
\hat{a}^{\dagger}},
\end{eqnarray}
where
\begin{eqnarray}
i \frac{d}{dt}\pmatrix{ u (t) \cr v(t)} + \pmatrix{\Omega^{(D)}
(t) & - \Omega^{(O)} (t) \cr \Omega^{(O)} (t) & - \Omega^{(D)}
(t)} \pmatrix{ u (t) \cr v(t)} = 0.
\end{eqnarray}
And we further impose the standard commutation relation
\begin{equation}
[ \hat{A} (t), \hat{A}^{\dagger} (t) ] = 1,
\end{equation}
which leads to the unimodular matrix
\begin{equation}
{\bf S} (t) = \pmatrix{u (t) & v(t) \cr v^*(t) & u^* (t)},~ {\rm
det} \Bigl({\bf S} \Bigr) = 1. \label{unim}
\end{equation}
The inverse transformation is given by
\begin{eqnarray}
\pmatrix{ \hat{a} \cr \hat{a}^{\dagger}} = \pmatrix{u^* (t) & -
v(t) \cr - v^*(t) & u (t)} \pmatrix{\hat{A} (t) \cr
\hat{A}^{\dagger} (t)}
\end{eqnarray}

With respect to the vacuum state (\ref{num st}) the position and
momentum dispersion relations are calculated to be
\begin{eqnarray}
\langle \hat{q}^2 \rangle_{(vac)} &=& \frac{1}{2 \omega_0} | u (t)
- v(t) |^2, \nonumber\\ \langle \hat{p}^2 \rangle_{(vac)} &=&
\frac{\omega_0}{2} | u(t) + v(t) |^2,
\end{eqnarray}
and the uncertainty relation to be
\begin{equation}
\langle \hat{q}^2 \rangle_{(vac)} \langle \hat{p}^2
\rangle_{(vac)} = \frac{1}{4} \Bigl\{1 - \Bigl(u (t) v^* (t) -
u^*(t) v (t) \Bigr)^2 \Bigr\}.
\end{equation}
By parameterizing Eq. (\ref{unim})
\begin{equation}
u(t) = \cosh r e^{i \theta_u},~ v(t) = \sinh r e^{i \theta_v},
\end{equation}
the uncertainty relation can be rewritten as
\begin{equation}
\langle \hat{q}^2 \rangle_{(vac)} \langle \hat{p}^2
\rangle_{(vac)} = \frac{1}{4} \Bigl\{1 + \sinh^2 r \sin^2
(\theta_u - \theta_v) \Bigr\}.
\end{equation}
Using the density operator (\ref{den op}), the thermal dispersion
function is found
\begin{equation}
\langle \hat{q}^2 \rangle_{(ther)} = \frac{1}{2 \omega_0}
\coth\Bigl(\frac{\beta \omega_0}{2}\Bigr) |u (t) - v(t)|^2.
\end{equation}

The second model is a system of coupled oscillators whose coupling
constants depend on time. The Hamiltonian is described by
\begin{equation}
\hat{H} (t) = \sum_{\alpha \beta} \Omega^{(D)}_{\alpha \beta} (t)
\Bigl(\hat{a}^{\dagger}_{\alpha} \hat{a}_{\beta} + \frac{1}{2}
\delta_{\alpha \beta} \Bigr) + \Omega^{(O)}_{\alpha \beta} (t)
\frac{1}{2}\hat{a}^{\dagger}_{\alpha} \hat{a}^{\dagger}_{\beta} +
\Omega^{(O)*}_{\alpha \beta} (t) \frac{1}{2}\hat{a}_{\alpha}
\hat{a}_{\beta}, \label{os2}
\end{equation}
where $\Omega^*_{\alpha \beta} = \Omega_{\beta \alpha}$ to
guarantee the hermiticity of the Hamiltonian. As for the first
model we may find pairs of first order invariant operators
\begin{eqnarray}
\hat{A}_{\alpha} (t) = \sum_{\beta} u_{\alpha \beta} (t)
\hat{a}_{\beta} + v_{\alpha \beta} (t) \hat{a}^{\dagger}_{\beta},
\nonumber\\ \hat{A}^{\dagger}_{\alpha} (t) = \sum_{\beta}
v_{\alpha \beta}^* (t) \hat{a}_{\beta} + u_{\alpha \beta}^* (t)
\hat{a}^{\dagger}_{\beta},
\end{eqnarray}
It is straightforward to see that Eq. (\ref{ln eq}) leads to the
system of equations
\begin{eqnarray}
&& i \dot{u}_{\alpha \beta} (t) + \sum_{\gamma}
\Omega^{(D)}_{\gamma \beta} (t) u_{\alpha \gamma} - \sum_{\gamma}
\Bigl(\Omega^{(O)*}_{\gamma \beta} (t) + \Omega^{(O)*}_{\beta
\gamma} (t) \Bigr) v_{\alpha \gamma} (t) = 0, \nonumber\\ && i
\dot{v}_{\alpha \beta} (t) - \sum_{\gamma} \Omega^{(D)}_{\gamma
\beta} (t) v_{\alpha \gamma} + \sum_{\gamma}
\Bigl(\Omega^{(O)}_{\gamma \beta} (t) + \Omega^{(O)}_{\beta
\gamma} (t) \Bigr) u_{\alpha \gamma} (t) = 0.
\end{eqnarray}
The commutation relations
\begin{equation}
[ \hat{A}_{\alpha} (t), \hat{A}^{\dagger}_{\beta} (t) ] =
\delta_{\alpha \beta},
\end{equation}
lead to the conditions
\begin{equation}
\sum_{\gamma} \Bigl(u_{\alpha \gamma} (t) u^*_{\beta \gamma} (t) -
v_{\alpha \gamma} (t) v^*_{\beta \gamma} (t) \Bigr) =
\delta_{\alpha \beta}.
\end{equation}

The final model for the open system is a system oscillator coupled
to an environment described by the Hamiltonian
\begin{equation}
\hat{H} = \hat{H}_{(sys)} + \hat{H}_{(sys-env)},
\end{equation}
where the system and the interaction are
\begin{eqnarray}
\hat{H}_{(sys)} &=& \omega_0 \hat{a}^{\dagger} \hat{a},
\nonumber\\ \hat{H}_{(sys-env)} &=& \nu (t) \hat{a}^{\dagger} +
\nu^* (t) \hat{a}.
\end{eqnarray}
Here, $\nu (t)$ is a complex function and of particular interest
is a stochastic function. Extending the algebra by including the
identity operator $\hat{e}$, a pair of first order invariant
operators of the form are found
\begin{eqnarray}
\pmatrix{ \hat{A} (t) \cr \hat{A}^{\dagger} (t)} = \pmatrix{u (t)
& v(t) \cr v^*(t) & u^* (t)} \pmatrix{\hat{a} \cr
\hat{a}^{\dagger}} + \pmatrix{w (t) \hat{e} \cr w^* (t) \hat{e}}.
\end{eqnarray}
The LvN equation (\ref{ln eq}) is satisfied with the solutions
\begin{eqnarray}
&& u (t) = u_0 e^{i \omega_0 t},~~ v(t) = v_0 e^{-i \omega_0 t},
\nonumber\\ && w (t) = i \int_{0}^{t} dt \Bigl(u_0 \nu (t) e^{i
\omega_0 t} - v_0 \nu^* (t) e^{- i \omega_0 t}\Bigr).
\end{eqnarray}
The Fock space is similarly constructed as in Eq. (\ref{num st}).

Using the inverse transformation
\begin{eqnarray}
\pmatrix{ \hat{a} \cr \hat{a}^{\dagger}} = \pmatrix{u^* (t) & -
v(t) \cr - v^*(t) & u (t)} \pmatrix{\hat{A} (t) - w(t) \hat{e} \cr
\hat{A}^{\dagger} (t) - w^* (t) \hat{e}},
\end{eqnarray}
it is straightforward to obtain the thermal expectation value of
the system Hamiltonian
\begin{equation}
\langle \hat{H}_{(sys)} \rangle_{(ther)} = \omega_0 \Bigl(u_0^*
u_0 + v_0^* v_0 \Bigr) \frac{1}{e^{\beta \omega_0} - 1} + \omega_0
v_0^* v_0 + \omega_0 |u(t) w^* (t) - v(t) w(t) |^2.
\end{equation}
The stochastic force has the correlation functions
\begin{eqnarray}
&& \langle \nu(t) \rangle_{(\nu)} = \langle \nu^*(t)
\rangle_{(\nu)} = 0, \nonumber\\ && \langle \nu (t') \nu (t)
\rangle_{(\nu)} = \langle \nu^* (t') \nu^* (t) \rangle_{(\nu)} =
0, \nonumber\\ && \langle \nu^* (t') \nu (t) \rangle_{(\nu)} =
f(t' - t),
\end{eqnarray}
where the average is taken with respect to a probability
distribution of fluctuating parameters. For a delta-correlated
force
\begin{equation}
\langle \nu^* (t') \nu (t) \rangle_{(\nu)} = D \delta (t' - t),
\label{del f}
\end{equation}
the energy expectation value becomes
\begin{equation}
\langle \hat{H}_{(sys)} \rangle_{(ther)} = \omega_0 \Bigl(u_0^*
u_0 + v_0^* v_0 \Bigr) \frac{1}{e^{\beta \omega_0} - 1} + \omega_0
v_0^* v_0 + \omega_0 D t.
\end{equation}
Thus the expectation value increases in proportion to the time of
interaction with the external driving force (\ref{del f}).
Similarly, for a Gaussian-correlated force
\begin{equation}
\langle \nu^* (t') \nu (t) \rangle_{(\nu)} = D \gamma e^{ - \gamma
| t' - t |}, \label{gaus f}
\end{equation}
we obtain the energy expectation value
\begin{eqnarray}
\langle \hat{H}_{(sys)} \rangle_{(ther)} &=& \omega_0 \Bigl(u_0^*
u_0 + v_0^* v_0 \Bigr) \frac{1}{e^{\beta \omega_0}-1} + \omega_0
v_0^* v_0 - \frac{2 D \gamma (\gamma^2 - \omega_0^2)}{\gamma^2 +
\omega_0^2} \nonumber\\ &&+ \frac{2 D \gamma^2}{\gamma^2 +
\omega_0^2} t + \frac{2D \gamma e^{- \gamma t}}{\gamma^2 +
\omega_0^2} \Bigl\{ (\gamma^2 - \omega_0^2) \cos \omega_0 t - 2
\gamma \omega_0 \sin \omega_0 t \Bigr\} .
\end{eqnarray}
The expectation value has constant terms, a linearly increasing
term and oscillating but decaying terms.

We are able to get the well-known results for these models
\cite{lindenberg}. However, the LvN approach provides us with a
clear picture for the open quantum system by making use of all the
lessons from quantum mechanics and proves technically much simpler
and straightforward than other methods.

\section{Open Fermion System}

It was not until a quite recent time that the LvN approach has
been applied to fermion systems \cite{finelli,kim6}. The
description of fermion systems requires not only the annihilation
and creation operators $ \hat{b}, \hat{b}^{\dagger}$ for the
particle but also those $\hat{d}, \hat{d}^{\dagger}$ for the
antiparticle. These operators satisfy the standard anticommutation
relations
\begin{eqnarray}
&&\{\hat{b} , \hat{b}^{\dagger} \} =  \{\hat{d}, \hat{d}^{\dagger}
\} = 1, \nonumber\\ &&\{\hat{b} , \hat{b} \} = \{\hat{b}^{\dagger}
, \hat{b}^{\dagger} \} = \{\hat{d} , \hat{d} \} =
\{\hat{d}^{\dagger} , \hat{d}^{\dagger} \} = 0, \nonumber\\
&&\{\hat{b} , \hat{d} \} = \{\hat{b} , \hat{d}^{\dagger} \} =
\{\hat{b}^{\dagger} , \hat{d} \} = \{\hat{b^{\dagger}} ,
\hat{d}^{\dagger} \} = 0.
\end{eqnarray}
The interaction between the fermion system and an external source
requires the Grassmann variables $\hat{c}$ and $\hat{c}^{\dagger}$
that satisfy the anticommutation relations
\begin{eqnarray}
&&\{\hat{c} , \hat{c}^{\dagger} \} = 1, \{\hat{c} , \hat{c} \} =
\{\hat{c}^{\dagger} , \hat{c}^{\dagger} \} = 0, \nonumber\\
&&\{\hat{c} , \hat{b} \} = \{\hat{c} , \hat{b}^{\dagger} \} =
\{\hat{c} , \hat{d} \} = \{\hat{c} , \hat{d}^{\dagger} \} = 0,
\nonumber\\ &&\{\hat{c}^{\dagger} , \hat{b} \} =
\{\hat{c}^{\dagger} , \hat{b}^{\dagger} \} = \{\hat{c}^{\dagger} ,
\hat{d} \} = \{\hat{c}^{\dagger} , \hat{d}^{\dagger} \} = 0.
\end{eqnarray}

The first open fermion model is described by the Hamiltonian
\begin{equation}
\hat{H} (t) = \hat{H}_{(sys)} + \hat{H}_{(sys-env)} (t),
\end{equation}
where the system and the interaction are given by
\begin{eqnarray}
\hat{H}_{(sys)} &=& \omega_0 \hat{b}^{\dagger} \hat{b},
\nonumber\\ \hat{H}_{(sys-env)} (t) &=& \zeta (t) \hat{c}
\hat{b}^{\dagger} + \zeta^* (t) \hat{b} \hat{c}^{\dagger}.
\end{eqnarray}
This model shares all the essential characteristics of a Dirac
field interacting with an external source which is described by
the Lagrangian
\begin{equation}
{\cal  L} = \overline{\psi} \Bigl(i \gamma_k {\partial}_k + m
\Bigr) \psi + \overline{\psi} \xi(t) + \overline{\xi} (t) \psi.
\end{equation}

We may find a pair of invariant operators
\begin{eqnarray}
\pmatrix{ \hat{B} (t) \cr \hat{B}^{\dagger} (t)} = \pmatrix{u (t)
& v (t) \cr v^*(t) & u^*(t)} \pmatrix{\hat{b} \cr
\hat{b}^{\dagger}} + \pmatrix{w^{(-)} (t) \hat{c} + w^{(+)}
\hat{c}^{\dagger} \cr w^{(+)*} (t) \hat{c} + w^{(-)*}
\hat{c}^{\dagger}},
\end{eqnarray}
where
\begin{eqnarray}
&&\ddot{u} (t) - \Bigl(\frac{\dot{\zeta}^* (t)}{\zeta^* (t)} + i
\omega_0 \Bigr) \dot{u} + \Bigl(\zeta (t)\zeta^* (t)+ i \omega_0
\frac{\dot{\zeta}^* (t)}{\zeta^* (t)} \Bigr) u(t) = 0,
\label{f-eq1}\\ &&\ddot{v} (t) - \Bigl(\frac{\dot{\zeta}
(t)}{\zeta (t) } - i \omega_0 \Bigr) \dot{v} (t) + \Bigl(\zeta
(t)\zeta^* (t) - i \omega_0 \frac{\dot{\zeta} (t)}{\zeta (t)}
\Bigr) v(t) = 0, \label{f-eq2}\\ && i \dot{w}^{(-)}(t) - \zeta (t)
u (t) = 0, \label{f-eq3}\\ && i \dot{w}^{(+)}(t) + \zeta^* (t)
v(t) = 0. \label{f-eq4}
\end{eqnarray}
Since Eqs. (\ref{f-eq2}) and (\ref{f-eq4}) are the complex
conjugate of Eqs. (\ref{f-eq1}) and (\ref{f-eq3}), respectively,
solutions can be found of the form
\begin{equation}
v (t) = v_0^* u^*(t) ,~ w^{(+)} = v_0^* w^{(-)*} (t),
\end{equation}
where $v_0$ is a complex number. We further impose the
anticommutation relation for all times
\begin{equation}
\{\hat{B} (t), \hat{B}^{\dagger} (t) \} = 1,
\end{equation}
which leads to the condition
\begin{equation}
\Bigl( u^* (t) u (t) + w^{(-)*} (t)w^{(-)} (t)\Bigr) \Bigl( 1 +
v_0^* v_0\Bigr) = 1.
\end{equation}
Under the assumption that the external source provides a
stochastic force with the correlations functions
\begin{eqnarray}
\langle \zeta (t) \hat{c} \rangle_{(\zeta)} = \langle \zeta^* (t)
\hat{c}^{\dagger} \rangle_{(\zeta)} =0, \nonumber\\ \langle \zeta
(t') \hat{c} \zeta^* (t) \hat{c}^{\dagger} \rangle_{(\zeta)} =
f(t' - t),
\end{eqnarray}
we find the expectation value of the Hamiltonian with respect to
the initial thermal equilibrium
\begin{eqnarray}
\langle \hat{H}_{(sys)} \rangle_{(ther)} = \frac{1}{u^* (t) u(t)
(1 - v_0^* v_0)} \Biggl[\frac{1}{e^{\beta \omega_0} + 1} +
\frac{v_0^* v_0}{1 - v_0^* v_0} \Bigl(1 + 2 w^{(-)*} (t)
w^{(-)}(t) \Bigr) \nonumber\\ + \int_{0}^{t} dt' \int_{0}^{t} dt''
f(t' - t'') u^*(t') u(t'') \Biggr].
\end{eqnarray}

The next open fermion model is motivated by a Dirac field
interacting with a scalar field
\begin{equation}
{\cal  L} = \overline{\psi} \Bigl(i \gamma_k {\partial}_k + m
\Bigr) \psi + \phi (t) \overline{\psi} \psi.
\end{equation}
Again without loss of generality we confine our attention to the
single-mode quadratic Hamiltonian
\begin{equation}
\hat{H} (t) = \hat{H}_{(sys)} (t) + \hat{H}_{(sys-env)} (t),
\label{f-ham}
\end{equation}
where the system and the system-environment Hamiltonians are,
respectively,
\begin{eqnarray}
\hat{H}_{(sys)} (t) &=& \Omega^{(D)} (t) \Bigl(\hat{b}^{\dagger}
\hat{b} - \hat{d} \hat{d}^{\dagger} \Bigr),\label{f-sys}\\
\hat{H}_{(sys-env)} (t) &=&  \Omega^{(O_c)} (t) \hat{b}^{\dagger}
\hat{d}^{\dagger} - \Omega^{(O_c)*} (t) \hat{b} \hat{d}  +
\Omega^{(O_e)} (t) \hat{b}^{\dagger} \hat{d} - \Omega^{(O_e)*} (t)
\hat{b} \hat{d}^{\dagger}. \label{f-sysenv}
\end{eqnarray}
We assume $\Omega^{(D)*} = \Omega^{(D)}$, so that the Hamiltonian
(\ref{f-ham}) becomes the unitary operator and its evolution
preserves the unitarity.

In Ref. \cite{kim6} the following type of invariant operators are
introduced
\begin{eqnarray}
 \hat{B} (t) &=& u_B^- (t) \hat{b} + u_B^+ (t) \hat{b}^{\dagger}
+ z_B^- (t) \hat{d} + z_B^+ (t) \hat{d}^{\dagger}, \nonumber\\
\hat{D} (t) &=& u_D^- (t) \hat{b} + u_D^+ (t) \hat{b}^{\dagger} +
z_D^- (t) \hat{d} + z_D^+ (t) \hat{d}^{\dagger} (t). \label{f-inv}
\end{eqnarray}
In fact, the algebra with the operators in Eq. (\ref{f-inv}) is
closed under the commutation relation because of the identity
between the commutator and the anticommutators
\begin{equation}
[ \hat{\cal O}_1, \hat{\cal O}_2 \hat{\cal O}_3] = \{\hat{\cal
O}_1, \hat{\cal O}_2 \}\hat{\cal O}_3 - \hat{\cal O}_2 \{\hat{\cal
O}_1, \hat{\cal O}_3 \}.
\end{equation}
The LvN equation leads to the following vector equation
\begin{eqnarray}
i \frac{d}{dt} \pmatrix{u^- (t) \cr u^+ (t)\cr z^-(t) \cr z^+(t)}
+ \pmatrix{\Omega^{(D)}(t)& 0& \Omega^{(O_e)*} (t)&
\Omega^{(O_c)*} (t)\cr 0 &- \Omega^{(D)} (t)& - \Omega^{(O_c)}(t)
& \Omega^{(O_e)} (t) \cr \Omega^{(O_e)} (t)&- \Omega^{(O_c)^*} (t)
& \Omega^{(D)} (t) & 0 \cr \Omega^{(O_c)} (t) & - \Omega^{(O_e)^*}
(t) & 0 & - \Omega^{(D)}(t)} \pmatrix{u^- (t) \cr u^+(t) \cr z^-
(t)\cr z^+(t)} = 0. \label{f-mat}
\end{eqnarray}
To simplify Eq. (\ref{f-mat}) two column vectors are introduced
\begin{equation}
U (t) = \frac{1}{\sqrt{2}}\pmatrix{u^- (t) + u^+(t) \cr\\ u^- (t)
- u^+ (t)},~~ Z (t) = \frac{1}{\sqrt{2}}\pmatrix{z^- (t) + z^+ (t)
\cr\\ z^- (t) - z^+(t)},
\end{equation}
in terms of which it can be rewritten as
\begin{eqnarray}
i \frac{d}{dt} \pmatrix{U (t) \cr Z(t)} + \pmatrix{\Omega^{(D)}
(t) \sigma_1 & M_F (t) \cr M_G  (t) & \Omega^{(D)} (t) \sigma_1}
\pmatrix{U (t) \cr Z (t)} = 0. \label{f-mat2}
\end{eqnarray}
Here,
\begin{eqnarray}
M_{F} (t) &=&  \Delta^{(0)} (t) I + \Delta^{(1)} (t)
\sigma_{1}+\Delta^{(2)} (t) \sigma_{2}+\Delta^{(3)} (t)
\sigma_{3}, \nonumber\\ M_{G} (t) &=&  - \Delta^{(0)} (t) I +
\Delta^{(1)} (t) \sigma_{1} - \Delta^{(2)} (t) \sigma_{2} -
\Delta^{(3)} (t) \sigma_{3},
\end{eqnarray}
where $\sigma_{i}$ are Pauli spin matrices and
\begin{eqnarray}
\Delta^{(0)} (t) = - \frac{1}{2} \Bigl( \Omega^{(O_e)} (t)
 - \Omega^{(O_e)*} (t) \Bigr),~~
\Delta^{(1)}(t) =  \frac{1}{2} \Bigl( \Omega^{(O_e)} (t) +
\Omega^{(O_e)*} (t) \Bigr), \nonumber\\ \Delta^{(2)} (t) = -
\frac{i}{2} \Bigl(\Omega^{(O_c)} (t) + \Omega^{(O_c)*} (t)
\Bigr),~~ \Delta^{(3)} (t) = \frac{1}{2} \Bigl(\Omega^{(O_c)} (t)
- \Omega^{(O_c)*} (t) \Bigr).
\end{eqnarray}

Further, it was shown in Ref. \cite{kim6} that the anticommutation
relations are consistently satisfied at all times
\begin{eqnarray}
&&\{\hat{B}(t), \hat{B}^{\dagger}(t) \} = \{\hat{D}(t),
\hat{D}^{\dagger}(t) \} = 1, \nonumber\\
 &&\{\hat{B}(t), \hat{B}(t)\} = \{\hat{B}^{\dagger}(t), \hat{B}^{\dagger}(t)\} =
\{\hat{D} (t), \hat{D}(t)\} = \{\hat{D}^{\dagger}(t),
\hat{D}^{\dagger}(t)\} = 0, \label{ant com}
\end{eqnarray}
Now, $\hat{B}_{\alpha}(t), \hat{B}^{\dagger}_{\alpha}(t)$ and
$\hat{D}_{\alpha}(t), \hat{D}^{\dagger}_{\alpha}(t)$ play the role
of the annihilation and creation operators for time-dependent
fermion system. And the number operators for the particle and
antiparticle are defined by
\begin{eqnarray}
\hat{N}_B (t) =  \hat{B}^{\dagger} (t) \hat{B} (t), \nonumber\\
\hat{N}_D (t) = \hat{D}^{\dagger} (t) \hat{D} (t),
\end{eqnarray}
whose number states provide the exact quantum states according to
Eq. (\ref{lin sup}):
\begin{eqnarray}
\hat{N}_B (t) \vert n_B, t \rangle = n_B \vert n_B, t \rangle,
\nonumber\\ \hat{N}_D (t) \vert n_D, t \rangle = n_D \vert n_D, t
\rangle,
\end{eqnarray}
where $n_B, n_D = 0, 1$.

\section{Conclusion}

In this paper we have applied the recently introduced
Liouville-von Neumann approach to open quantum systems of bosons
and fermions. It is based on physically well-grounded assumptions.
Firstly, the quantum evolution of all systems, either
time-independent or time-dependent, obeys the Tomonaga-Schwinger
equation or the functional Schr\"{o}dinger equation. Secondly, the
statistical nature of systems is determined by the density
operator. Technically the Liouville-von Neumann equation is solved
to yield invariant operators which determine not only the Hilbert
space of exact quantum states but also the density operators.
Therefore, the Liouville-von Neumann approach describes the
nonequilibrium quantum evolution of open systems. Moreover, the
Liouville-von Neumann approach can also be applied to open fermion
systems. In summary, the Liouville-von Neumann approach is a
unified method that can be applied both to the equilibrium quantum
evolution of time-independent systems and to the nonequilibrium
quantum evolution of time-dependent boson and fermion systems.

\vspace{1cm}

The author would like to thank F.C. Khanna, K.-S. Soh and J.H. Yee
for collaborations and useful discussions. This work was supported
by KRF under Grant No. 1998-001-D00354.

\end{document}